\documentstyle[aps,preprint]{revtex}
\input epsf.tex
\def\DESepsf(#1 width #2){\epsfxsize=#2 \epsfbox{#1}}
\begin{document}

\preprint{\vbox{\hbox{UH511-880-97}}}
\draft
\title{Up-Down Asymmetry:\\A Diagnostic for Neutrino Oscillations}
\author{John~W.~Flanagan, John~G.~Learned, and Sandip~Pakvasa}
\address{Department of Physics and Astronomy, University of Hawaii\\
Honolulu, HI 96822}
\date{23 September 1997}
\maketitle

\begin{abstract}

We propose a simple test of underground neutrino data to
discriminate amongst neutrino oscillation models. It uses the
asymmetry between downward-going events and upward-going events, for
electron and muon events separately. Because of the symmetry of
typical underground detectors, an asymmetry can be compared with
calculations with little need for the intermediary of a simulation
program.  Furthermore, we show that the various oscillation
scenarios give rise to dramatically differing trajectories of
asymmetry versus energy for muons and electrons.  This permits a
clean distinction to be drawn between models.

\end{abstract}

\section{Introduction}

The atmospheric neutrino anomaly has been with us now for more than
ten years.  It was discovered serendipitously in the largest
detectors built for nucleon decay searches, IMB\cite{imb_atmanomaly}
and Kamioka\cite{kam_atmanomaly}, and confirmed recently in
Soudan\cite{soudan_atmanomaly}. It appeared as a deficit in the
ratio of muon-like to electron-like neutrino interactions within the
fiducial volumes of these massive Cherenkov detectors.  The
atmospheric neutrinos in the range of $1~GeV$ of energy originate
high in the atmosphere from the decay of pions and muons. The ratio
of the $\nu_\mu$ flux to $\nu_e$ flux is thus closely constrained by
well known decay kinematics to be in the ratio of 2:1 (actually,
2.2:1).  The experiments have found that the apparent ratio of
observed flux flavor ratio to expected flux flavor ratio is closer
to $R=0.60\pm 0.05$.  The result is usually presented in terms of
this ratio of ratios because the absolute neutrino fluxes are not
very well known ($\sim20\%$, or perhaps even worse), whereas the
$\phi(\nu_e)/\phi(\nu_\mu)$ is known to several percent.
The simplest explanation for the anomaly seems to be the one
originally suggested, namely neutrino oscillations\cite{learned88}.

Until the present time the size of the underground detectors has
limited the number of events to a few hundred, and the energy of
those events to less than about $1.5~GeV$. Previous data have not
shown conclusive zenith angle dependence as expected from neutrino
oscillations.  At energies of a few GeV however, we must begin to
observe angular effects, if neutrino oscillations are the cause of
the anomaly.  Indeed, the first public presentations of data from
the new massive Super-Kamiokande experiment do seem to indicate some
angular variation\cite{totsuka97}.

Our purpose in this note is to set the stage to interpret future
data in terms of discriminating amongst the many scenarios which
have been constructed to explain one or more of the three
outstanding ``problems'' in neutrinos: the solar neutrino
problem\cite{solarnu_summary}, the LSND effect\cite{lsnd97}, and the
presently considered atmospheric neutrino anomaly.

We define an asymmetry in direction as simply
\begin{equation}
A = \frac{D-U}{D+U}
\end{equation}
where $D$ are the number of downward-going events and $U$ are
upward-going events, for each of muon neutrino events and electron
neutrino events.  We assume the detector to be up/down symmetric,
and the data set to be free of significant (presumably
downward-going) background and crossover between muon and electron
type of charged-current events.

We ignore the effect of $\nu_\tau$ in our calculations, as well as
the contamination of $\nu_e$ events by neutral current interactions.
The $\nu_\tau$ charged-current cross section is sufficiently small
at the energies discussed herein that it makes a negligible
contribution (to both muon and electron events).  Thus for present
considerations, oscillations between $\nu_\mu$ and $\nu_\tau$ are
indistinguishable from oscillations between $\nu_\mu$ and a new
sterile neutrino species.  Neutral currents are a small fraction of
events, and in any case should show no asymmetry (except in the
sterile neutrino case).

\section{Calculations}

We now calculate this asymmetry for a wide variety of oscillation
scenarios.  We consider the following cases:  

(A) two-flavor mixing (a) $\nu_\mu - \nu_\tau$ and (b) $\nu_\mu
-\nu_e$;

(B) three-flavor mixing with a variety of choices for mass-mixing
parameters;

(C) sterile maximal mixing; and 

(D) massless neutrino mixing.

It will be shown that these various scenarios lead to very different
predictions in the sign, magnitude or the energy dependence of the
asymmetry and hence are (relatively) easy to distinguish from one
another once there is sufficient data available.

\subsection{Two-Flavor Mixing}

\paragraph{$\nu_\mu - \nu_\tau$ Mixing.}

This is the simplest case possible\cite{learned88}. The $\nu_e$ flux
is unaffected and the $\nu_\mu$ flux modified as
\begin{eqnarray}
N_\mu &=& N^0_\mu (P_{\mu \mu})  \nonumber \\
\mbox{where} \ \ P_{\mu \mu} &=& 1 - \sin^2 2 \theta \sin^2 
\frac{\delta m^2 L}{4E} .
\end{eqnarray}

Hence $A_e$ is essentially zero and independent of energy.  At low
enough energies $P_{\mu \mu} \approx 1- \frac{1}{2} \sin^2 2 \theta$
and is independent of L and hence $A_\mu \rightarrow 0$ at low
energies.  At high energies; L/E is negligible for down $\nu'_\mu s$
and hence
\begin{equation}
N_\mu^d = N^0_\mu.
\end{equation}

For upward-going $\nu_\mu's \ P^u_{\mu \mu} \approx (1 -\frac{1}{2}
\sin^2 2 \theta)$ and $N_\mu^u \approx (1-\frac{1}{2} \sin^2 2
\theta) N_\mu^0$ and hence:
\begin{equation}
A_\mu  = 
\frac{1/2 \sin^22 \theta}{1 + 1/2 \sin^2 2 \theta}
\end{equation}
$A_\mu$ has a maximum of 1/3 when $\sin^2 2 \theta = 1$.  Note that
at high enough energies $A_\mu$ will come back asymptotically to
zero.  

\paragraph{$\nu_\mu - \nu_e$ Mixing}

In this case\cite{learned88}
\begin{eqnarray}
N_{\mu} &=& N^0_\mu (P \ + r (1-P)) \nonumber \\
N_e    &=& N^0_e (P  + (1/r) (1-P))
\end{eqnarray}
where $P = P_{\mu \mu} = P_{ee}$, as in Equation (2), and 
$r = N^0_e / N^0_\mu $.
Again at low energies $A_e = A_\mu \approx 0$. At high energies
\begin{eqnarray}
N^d_\mu &=& N^0_\mu,\  N_\mu^u = N^0_\mu \ (P + r(1-P)) \nonumber\\
N_e^d &=& N^0_e, \ N_e^u = N^0_e \ (P + (1/r)(1-P)) \nonumber \\
A_\mu & = &
  \frac{(1-P) - r (1-P)}{(1+P) + r (1-P)} \\
A_e &=&
\frac{(1-P) - 1/r (1-P)}{(1+P) + 1/r (1-P)} \nonumber
\end{eqnarray}

For $P=1/2$ we get the limiting values:
\begin{equation}
A_\mu \ = \frac{1-r}{3+r} 
\end{equation}
and
\begin{equation}
A_e = -\frac{1-r}{3r+1} .
\end{equation}

Recall that $r\sim 0.45$ at low energies, decreases to $0.3$ at 
$E_\nu \sim 5~GeV$ and eventually becomes negligible. Note that
$A_\mu$ and $A_e$ always have {\it opposite} signs in this case.

\subsection{Three-Flavor Mixing}

There are two ways to account for all three neutrino anomalies
(solar, atmospheric and LSND) with just three flavors.  In one, due
to Cardall and Fuller\cite{cardall96}, a single $\delta m^2$ is
expected to account for both the atmospheric low energy anomaly as
well as the LSND observations.  In particular, $0.3~eV^2 \sim \delta
m^2_{31}$ $ \sim \delta m^2_{32} >> \delta m^2_{12} \sim 10^{-5}~eV^2$,
with large $\nu_\mu - \nu_\tau$ mixing.  In this case the resulting
probabilities are very similar to the two-flavor $\nu_\mu - \nu_\tau$
mixing, but with a large $\delta m^2$ of $0.3~eV^2$.  As a result of
the large $\delta m^2$ very little zenith angle dependence or
asymmetry (neither $A_e$ nor $A_\mu$) is expected.

The other three-flavor solution, due to Acker and
Pakvasa\cite{acker97}, accounts for both solar and atmospheric
anomalies with a single $\delta m^2$ and with large mixing between
$\nu_e$ and $\nu_\mu$, the mass pattern being $2~eV^2 \sim \delta
m^2_{31}$ $\sim \delta m^2_{32} >> \delta m^2_{12} \sim 5 \times
10^{-3}~eV^2$.  The probabilities in this case are essentially
identical to the two-flavor $\nu_\mu - \nu_e$ case with large
mixing.  In both of these scenarios, it is possible to have nearly
degenerate neutrinos with cosmologically significant total mass.

There are also other scenarios with three-neutrino mixing with a
wide range of mixing patterns{\cite{fogli96}.  In general, we expect
them to yield asymmetries which will interpolate between the two
limiting cases of $\nu_\mu - \nu_\tau$ and $\nu_\mu - \nu_e$ mixing.

We consider as a unique and interesting example, the maximal mixing
proposal of Harrison, Perkins and Scott\cite{harrison97}.  The
assumption is that $\delta m_{32}^2 \sim \delta m_{31}^2 >> \delta
m^2_{12} \sim 10^{-11} eV^2$ and $\delta m_{31}^2$ is in the range
of $10^{-2} - 10^{-3}~eV^2$. With the assumed maximal mixing, the
probabilities for $L/E$ in the appropriate atmospheric range are
given by
\begin{eqnarray}
P_{\mu \mu} &=& P_{ee}  = 1- 8/9 \ \sin^2
  \frac{\delta m_{31}^2 L}{4E}    \nonumber \\
P_{\mu e} &=& P_{e \mu} ( = P_{e \tau} = P_{\mu \tau}) = 4/9
\sin^2  \frac{\delta m_{31}^2 L}{4E}
\end{eqnarray}
The expected asymmetries can be easily written down
\begin{equation}
A_\mu=
 \frac{(1-P_{\mu \mu}) - r P_{\mu e}} {(1-P_{\mu \mu}) + r P_{\mu e}},
A_e = \frac{(1-P_{ee}) - 1/r P_{e\mu }} {(1+P_{ee}) - 1/r P_{e \mu }}
\end{equation}
Hence both $A_\mu$ and $A_e$ are small at low energies and at high
energies $P^d_{\mu\mu} =P^d_{ee} = 1$ and $P^d_{e \mu} = 0$, 
whereas $P^u_{\mu\mu} =P^u_{ee} = 5/9$ and $P^u_{e \mu} = 2/9$.
And so
\begin{eqnarray}
A_\mu &=&
\frac{4/9 - r \ 2/9}{14/9 + r \ 2/9} = \ \frac{2-r}{7+r}  \nonumber
\\
\mbox{and} \ \
A_e &=&
\frac{4/9 - 1/r^(2/9)}{14/9 + 1/r (2/9)} = \ -\frac{1-2r}{7r+1} .
\end{eqnarray}

\subsection{Sterile Maximal Mixing}

In the scheme of Foot and Volkas\cite{foot95}, $\nu_\mu$ mixing
maximally with a new sterile $\nu_{\mu'}$ with a $\delta
m^2_{\mu\mu'}\sim 5 \times 10^{-3}~eV^2$ accounts for the low-energy
atmospheric neutrino anomaly.  Hence, the muon asymmetry is
identical to the one in the case of $\nu_\mu - \nu_\tau$
oscillations, as discussed above.  In addition, $\nu_e$ mixes
maximally with a sterile $\nu_{e'}$, and when $\delta m_{ee'}^2$ is
in the range of $10^{-3} eV^2$ electron-neutrinos will also
oscillate and get depleted.  The resulting electron asymmetry is
strikingly different from the case of $\nu_\mu - \nu_e$ oscillations
in having the opposite sign, which makes it unique and easy to
distinguish.  $A_e$ and $A_\mu$ will differ only in slightly
different energy (or L/E) dependence but be otherwise similar and
always have the {\it same} sign.

\subsection{Massless Neutrinos}

If neutrinos are massless, there can still be mixing and
oscillations.  Two possibilities have been considered in the
literature.  One is the case where different flavors couple
differently to gravity\cite{gasperini88} and the other is a
breakdown of Lorentz invariance where each particle may have its own
maximum speed\cite{glashow97}.  The oscillation phenomenology is
identical for both cases.  The survival probability in the
two-flavor limit is given by
\begin{equation}
P_{\mu\mu} = P_{ee} = 1 - \sin^2 2 \theta \sin^2
\left ( \frac{1}{2} \delta v\ \ EL \right )
\end{equation}
where $\theta$ is the flavor mixing angle, and $\delta v$ is the
small parameter characteristic of violation of equivalence principle
or Lorentz invariance.  Note the strikingly different dependence on
$L$ and $E$: $L\times E$ instead of $L/E$.  Remarkably, an allowed
choice of parameters is able to account for both solar and
atmospheric neutrinos\cite{pantaleone96}: $\sin^2 2 \theta_v \approx
0.8$ to 1 and $\frac{\delta v}{2} \sim \ \ 10^{-2}-10^{-3}(km
-GeV)^{-1}$.  The expressions are the same as in the $\nu_\mu-\nu_e$
case except that $\sin^2 \frac{\delta m^2 L}{4E}$ is replaced by
$\sin^2 (\delta v/2 L E)$.  As a result, the roles of low and high
energy are reversed.  The asymmetries $A_\mu$ and $A_e$ become
rather small at high energies; at {\it low} energies they are given
by Eq.(12).

\section{Numerical Results}

We have performed numerical calculations of the models
discussed above.  They are explicitly:
\begin{enumerate}
\item Simple two-flavor oscillations between $\nu_\mu$ and
$\nu_\tau$ \cite{learned88}. The example is for $\delta m^2 =
0.005~eV^2$, $\sin^2 2\theta = 1$.
\item Two-flavor oscillations between $\nu_\mu$ and $\nu_e$ with the
same parameters as above.  (The Acker-Pakvasa\cite{acker97} scheme
leads to the same result).
\item Three-flavor mixing {\it a' la} Cardall-Fuller\cite{cardall96}.
\item Three-flavor maximal mixing scheme of
Harrison-Perkins-Scott\cite{harrison97}.
\item Sterile maximal mixing of Foot-Volkas\cite{foot95}.
\item Massless neutrino mixing, where we take $\delta v/2 \sim
10^{-3}~(km-GeV)^{-1}$\cite{pantaleone96}.
\end{enumerate}

In Figure \ref{fig:asym} we show results for $A_\mu$ and $A_e$ as
functions of energy from more detailed calculations. We calculated
energy spectra between 0.2 and 5.0 GeV for a detector with an
exposure of 22 kiloton-years (approximately one year of
Super-Kamiokande data).  We use the Bartol flux model, and a simple
quark model for the charged-current cross section, and assume a
perfect detector\cite{stanev96}.  Detailed calculations for a
particular instrument will of course vary, but the asymmetry will
change little, the general behavior illustrated being insensitive
to the details.

We show the trajectories of $A_\mu$ versus $A_e$ in Figure
\ref{fig:muvse} for the six models.  Note that there are small
asymmetries at low energies, due to the inhomogeneity of the earth's
magnetic field, as incorporated into the atmospheric flux model we
employ\cite{stanev96}.  It is clear that with good statistics all
scenarios can be clearly distinguished by both energy dependence and
relative signs of $A_\mu$ and $A_e$.  In particular it is noteworthy
that the first model, currently seemingly favored in preliminary
reports from Super-Kamiokande\cite{totsuka97}, stands out distinctly
from all others.  It is straightforward to plot the expected
asymmetries in other scenarios or different choices of parameters.

\section{Conclusions}

In the foregoing we have presented a case for employing the
up-to-down asymmetry of neutrino interactions in underground
detectors as a discriminator for some of the many neutrino
oscillation schemes which have been discussed as solutions to
various combinations of the current three neutrino puzzles (solar,
atmospheric and LSND).  The asymmetry has the virtue that it can be
calculated directly from data (using only particle identification,
energy and direction), without aid of simulation programs.  It is
self-normalizing and independent of flux model calculations, and
tests electron and muon data separately.

\section*{Acknowledgments}

We want to thank many colleagues for discussions of the foregoing
analysis, in particular: Todor Stanev, Walt Simmons, Xerxes Tata,
Shigenobu Matsuno, many members of the Super-Kamiokande
Collaboration, and Kazu Mitake for providing a stimulating
environment.

\begin{figure}[htb]
\centerline{\epsfysize 6.0 truein \epsfbox{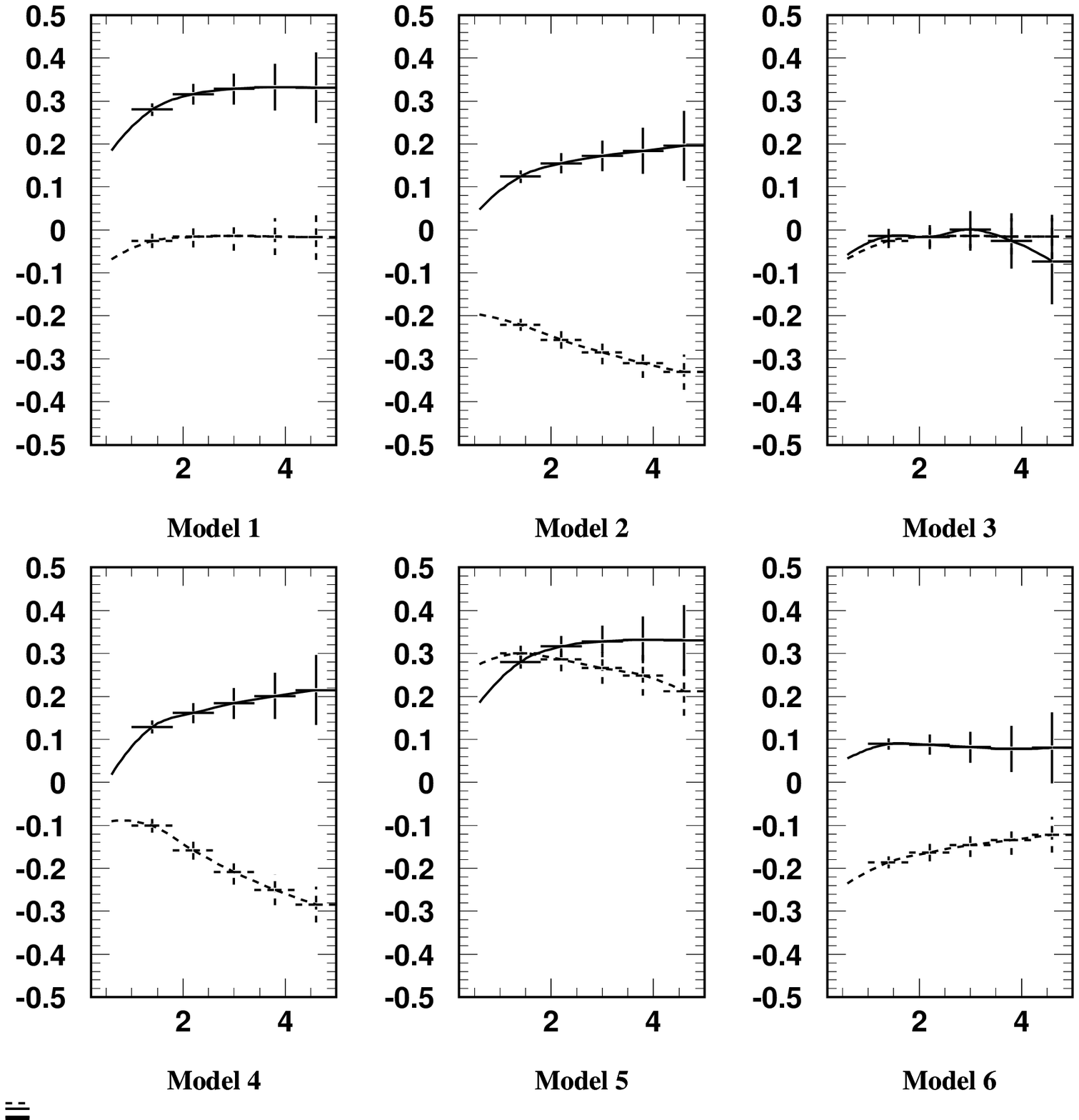}}
\caption{The muon (solid lines) and electron (dashed lines)
asymmetries versus energy (GeV), for the 6 oscillations models 
considered herein (see text).}
\label{fig:asym}
\end{figure}

\begin{figure}[htb]
\centerline{\epsfysize 6.0 truein 
\epsfbox{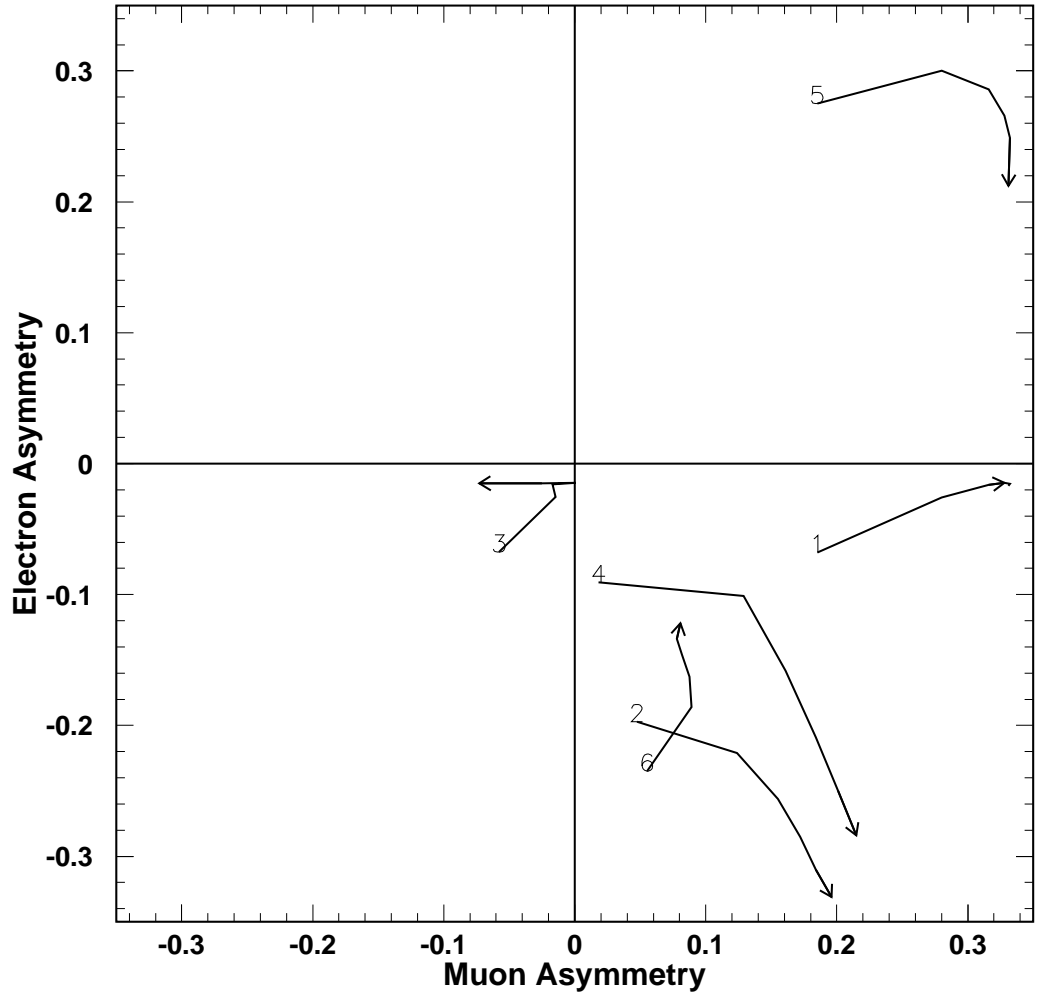}}
\caption{The trajectories of the muon asymmetry and electron
asymmetry for the 6 oscillation models considered herein. The
arrowheads point in the direction of increasing charged lepton
energy, which ranges from $0.2$ to $5.0~GeV$ in these calculations.}
\label{fig:muvse}
\end{figure}

\end{document}